\documentclass{aa}  

\usepackage{xcolor}
\usepackage{graphicx}
\usepackage[varg]{txfonts}
\usepackage{isotope}
\usepackage{siunitx}
\usepackage{ulem}
\usepackage{bm}
\usepackage{natbib}
\usepackage{url}
\usepackage{xspace}
\usepackage{placeins}
\usepackage{capt-of}
\usepackage{booktabs}
\usepackage{subcaption}  

\bibpunct{(}{)}{;}{a}{}{,}

\usepackage[colorlinks=true,citecolor=blue,linkcolor=magenta,urlcolor=blue]{hyperref}
\newcommand{\gaia}{\textit{Gaia}}
\newcommand{\gstarfull}{Gaia DR3 5446737753669901568}
\newcommand{\gstar}{J1022-3414}
\usepackage{orcidlink}

\newcommand{\msun}{$M_{\odot}$}
\newcommand{\rsun}{$R_{\odot}$}

\newcommand{\teff}{$T_\mathrm{eff}$}
\newcommand{\logg}{$\log g$}
\newcommand{\lp}{LP\,40$-$365}
\newcommand{\kms}{km\,s$^{-1}$}

\newcommand{\Ion}[2]{#1\,\textsc{#2}}
\begin{document} 

   \title{Discovery of a runaway star likely ejected by a Type Iax Supernova}

   \author{A. Bhat
          \inst{1}\thanks{Email: aakashbhat7@gmail.com}\orcidlink{0000-0002-4803-5902}
          \and
          M. Hollands
          \inst{2}
          \and
          M. Dorsch
          \inst{1}\orcidlink{0000-0001-5400-2368}
          \and
          S. Geier
          \inst{1}
          \and
          U. Heber
          \inst{3}
          \and
          D. Koester
          \inst{4}
          \and
           R. Pakmor\inst{5}\orcidlink{0000-0003-3308-2420}
            \and
          Ken J. Shen\inst{6}\orcidlink{0000-0002-9632-6106}
          }

   \institute{Institut für Physik und Astronomie, Universität Potsdam, Haus 28, Karl-Liebknecht-Str.\ 24/25, 14476 Potsdam, Germany
        \and
        Department of Physics,
        University of Warwick,
        Coventry CV4 7AL
        UK
         \and
             Dr.\ Karl\ Remeis-Observatory \& ECAP, Astronomical Institute, Friedrich-Alexander University Erlangen-Nuremberg, Sternwartstr.~7, 96049 Bamberg, Germany
             \and
             Institut für Theoretische Physik und Astrophysik, Christian-Albrechts-Universität zu Kiel, Leibnizstraße 15, 24118, Kiel, Germany
             \and
            Max Planck Institut für Astrophysik, Karl-Schwarzschild-Straße 1, 85748 Garching bei München, Germany
            \and
            Department of Astronomy and Theoretical Astrophysics Center, University of California, Berkeley, CA 94720-3411, USA
             }


 
  \abstract
   {Over the past decade, runaway stars have been identified, believed to originate either as surviving donors of Type Ia supernovae or as partially deflagrated accretors producing Type Iax supernovae. 
   While the former have been extensively studied recently, the origins of the latter (also called \lp\ type stars) remain under-explored and therefore less well understood. So far seven such objects are known. In this paper, we report the discovery of a new \lp\ type runaway star, notably hotter than previously studied members of this class. Spectral analysis confirms that its atmosphere is neon- and oxygen-dominated, consistent with earlier analyses of other \lp\ type stars. Kinematic analysis indicates that the star has a high probability of being unbound from the Galaxy and was most likely ejected from the Galactic disk approximately 2.8~Myr ago with an ejection velocity exceeding $600~\mathrm{km\,s^{-1}}$. This result further emphasizes the discrepancy between the abundance yields and kick velocities predicted by white dwarf deflagration models and those observed in stars of \lp\ type, underscoring the need for a reassessment of such systems.

 } 


   \maketitle
%

\section{Introduction}

Type Ia supernovae (SNe Ia) are widely interpreted as thermonuclear explosions of carbon-oxygen white dwarfs in binary systems \citep{1982ApJ...253..798N,2025A&ARv..33....1R}. These explosions are of major importance in cosmology because their peak luminosities can be calibrated (via empirical light-curve relations) to act as ``standardisable candles'' for measuring extragalactic distances and the Hubble constant \citep{1985Natur.314..337A}. Their use has underpinned the discovery of the accelerating expansion of the universe \citep{1998AJ....116.1009R} and plays a central role in current discussions on the Hubble tension \citep{2024ApJ...973L..14D}. The association between Type Ia SNe and exploding white dwarfs is supported by multiple lines of observational evidence. Several SNe have been directly linked to degenerate progenitors \citep{2011Natur.480..344N}, and pre-explosion imaging has further constrained progenitor systems by ruling out luminous non-degenerate companions in nearby events \citep{2012ApJ...744L..17B,2014ApJ...790....3K}.

More recently, the discovery of multiple populations of runaway stars has provided an additional line of evidence linking white dwarfs in binary systems to supernova progenitor systems. So far eight stars have been observed \citep{shen2018,kareemfast,hollands2025} which travel faster than $1000$ \kms\ through our Galaxy (and so are also called hypervelocity stars). The extreme velocities and distinctive spectral abundances of these stars can be explained only within the framework of double white dwarf binaries, where the compact nature of the components allows for the exceptionally high orbital speeds required, which are unattainable in other systems. For the last few years, it was believed that these objects had been produced by the so-called dynamically-driven, double-degenerate, double detonation (D$^6$) mechanism. In this scenario, a white dwarf transfers mass to a more massive white dwarf through dynamically unstable Roche-lobe overflow \citep{2007A&A...476.1133F, Guillochon,2010ApJ...719.1067K,Dan2011,Dan2012,2013ApJ...770L...8P,Shen2018a,2021ApJ...919..126B,pakmor2022}. The more massive white dwarf explodes in a Type Ia supernova and the secondary is ejected as a hypervelocity star. Although these theories were successful in predicting a surviving donor star, they were unable to explain the inflated nature of the observed stars \citep{Bhat1}. Recent work has instead shown that these stars can be produced in violent mergers of CO or HeCO white dwarfs \citep{glanz1,pakmor2025,Bhat2}. These scenarios can produce lower mass stars which are faster than those produced through the D$^6$ scenario due to the donor white dwarf plunging into the accretor and being disrupted already at the moment of explosion.

In connection with Type Ia SNe, another set of events has been discovered to exhibit peculiar light-curve and spectral properties that could not be reconciled with existing models or observations of normal Type Ia explosions \citep{2003PASP..115..453L}. These SNe are characterized by lower ejecta velocities ($<8000~\mathrm{km\,s^{-1}}$) and the absence of late-time signatures from iron-group elements. To account for these anomalies, pure deflagration models involving near-Chandrasekhar-mass white dwarfs were proposed \citep{2006AJ....132..189J,2007PASP..119..360P}, and a  new subclass known as Type Iax SNe was identified \citep{Foley2013}. In some of these models deflagrations were unable to unbind the whole star, a fast-moving runaway remnant was predicted \citep{2012ApJ...761L..23J}, and the prototype star \lp\ was later on found by \citet{2017Sci...357..680V}. Similar to the case of Type Ia progenitors, a population of six neon- and oxygen-rich high-velocity ($>500~\mathrm{km\,s^{-1}}$) runaway stars with masses in the range $0.1-0.4$ \msun\ have since been found, providing direct evidence for white-dwarf progenitors of Type Iax SNe \citep{2017Sci...357..680V,raddi2017,raddi2018,raddi2019,kareemfast}. The elemental abundances and kinematics of these stars are best explained if their progenitor white dwarfs underwent partial thermonuclear explosions, leaving behind bound remnants. These objects, known as LP 40-365 stars, are named after the prototype discovered by \citet{2017Sci...357..680V}. Unlike the canonical Type Ia scenario, in which the donor star is ejected, the runaway in this case is thought to be the former accretor \citep{2012ApJ...761L..23J,2022A&A...658A.179L,2024MNRAS.532.1087M}.

The precise conditions determining whether a white-dwarf explosion results in a partial or complete disruption remain uncertain. Similarly, the nature and role of the donor star have not yet been fully constrained, although progenitor systems leading to Type Iax SNe are generally believed to consist of a white dwarf accreting from a helium-rich companion \citep{2020MNRAS.493..986T}. Recent studies suggest that, for accretion-induced explosions in carbon-oxygen (CO) white dwarfs with non-degenerate donors, white dwarfs with masses below $1.365\,M_\odot$ always undergo partial deflagrations \citep{2025arXiv250716907M}. The hypothesis that CO white dwarfs serve as the main progenitors of deflagrations has been investigated \citep{2012ApJ...761L..23J,2022A&A...658A.179L}, but the observed surface abundances of the surviving stars do not match the theoretical yields \citep{raddi2019}. The observed $^{20}$Ne abundance is in particular much higher than the theoretical predictions. More recent work instead models the observed stars as deflagrations in oxygen-neon (ONe) white dwarfs \citep{ken2025}, although the predicted abundances of these elements in ONe white dwarfs are still too low \citep{jones2016,2019A&A...622A..74J}. Furthermore, the kicks produced in both CO and ONe WD deflagrations are not consistent with the observed runaway velocities (maximum $\sim 300~\mathrm{km\,s^{-1}}$, for ONe white dwarfs as reported by \citealt{2022A&A...658A.179L}). Consequently, no single scenario currently explains both the abundances and kinematics, but enlarging the sample of confirmed objects will help place stronger observational constraints on the progenitor systems.

In this paper, we detail the discovery and analysis of a new runaway star of \lp\ type, which we found serendipitously in a search for runaway subdwarfs. The paper is structured as follows: Section \ref{sect:data} describes the target selection and data acquisition. Sections \ref{sect:spec} and \ref{sect:sed} deal with the spectrophotometric properties. Section \ref{sect:kin} details the kinematic properties and Section \ref{sect:evol} describes the evolutionary status of the star. Finally, we discuss the results in the context of the broader context of Type Iax SNe and their progenitors.
  
\section{Target Selection and data acquisition}
\label{sect:data}

 \gstarfull\ or J102230.26-341420.5 (\gstar\ from here on) was identified in a search for hypervelocity stars undertaken at ESO VLT with the FOcal Reducer and low dispersion Spectrograph 2 (FORS2). We utilized the simulation data from \citet{neunteufel}, and used a modified query from \citet{kareemfast} to the \gaia\ DR3 catalogue \citep{2023A&A...674A...1G}. \citet{kareemfast} queried \gaia\, database for stars with $G_{\rm BP}-G_{\rm RP}<0.5$, and proper motion (pm) $>50$. Furthermore, the parallax uncertainties had to be consistent either with the stars beings far away (negative parallaxes) or with tangential velocities greater than $400$ \kms\,. This method led to the discovery of several hypervelocity white dwarfs. Since the fastest objects have already been studied, we limited ourselves to objects with pm between $20-50\ \rm mas\ yr^{-1}$. At distances $>8$ kpc (the distance of US 708), this pm corresponds to a tangential velocity $v_{\rm tan}>760$ \kms. This velocity lies in the predicted upper limit for hypervelocity subdwarfs by \citet{neunteufel}. As described in \citet{Uli2018}, the spectral energy distribution (SED) of a hot subdwarf star can be modelled using stellar atmospheric models. These models are fit to the SED using a $\chi^2$-minimization algorithm which considers the angular diameter of the star and interstellar reddening as free parameters. We modelled the candidates according to this method. This allowed us to pre-select hotter candidates and reject halo contaminants like blue horizontal branch (BHB) stars. The astrometric parameters are given in Table~\ref{tab:kinematics}. 

\begin{table}
\centering
\caption{Astrometric parameters of \gstar}
\begin{tabular}{lcc}
\toprule\toprule
Parameter & Units & Value \\
\midrule
Right ascension (RA) &  deg & $155.62617$ \\
Declination (Dec) & deg & $-34.23891$ \\
Gaia $G$ magnitude & mag & 18.99 \\
Parallax ($\varpi$) &  mas & 0.099 $\pm$ 0.230 \\
Proper motion in RA ($\mu^*_{\alpha}$) &  mas\,yr$^{-1}$ & $17.56 \pm 0.18$ \\
Proper motion in Dec ($\mu_{\delta}$)&  mas\,yr$^{-1}$ & $23.75 \pm 0.19$ \\
RUWE & - & 0.979 \\
\bottomrule
\end{tabular}
\tablefoot{
Astrometric parameters from Gaia DR3. The proper motion in right ascension, $\mu^*_{\alpha}$, includes the $\cos\delta$ factor. The renormalized unit weight error (RUWE), is a quality criterion to quantify deviations from  single star astrometric solutions. A value less than $1.4$ is normally considered to be consistent with a single-star fit \citep{gaia2}.
}
\label{tab:kinematics}
\end{table}

\begin{figure*}
    \centering
    \includegraphics[width=0.99\linewidth]{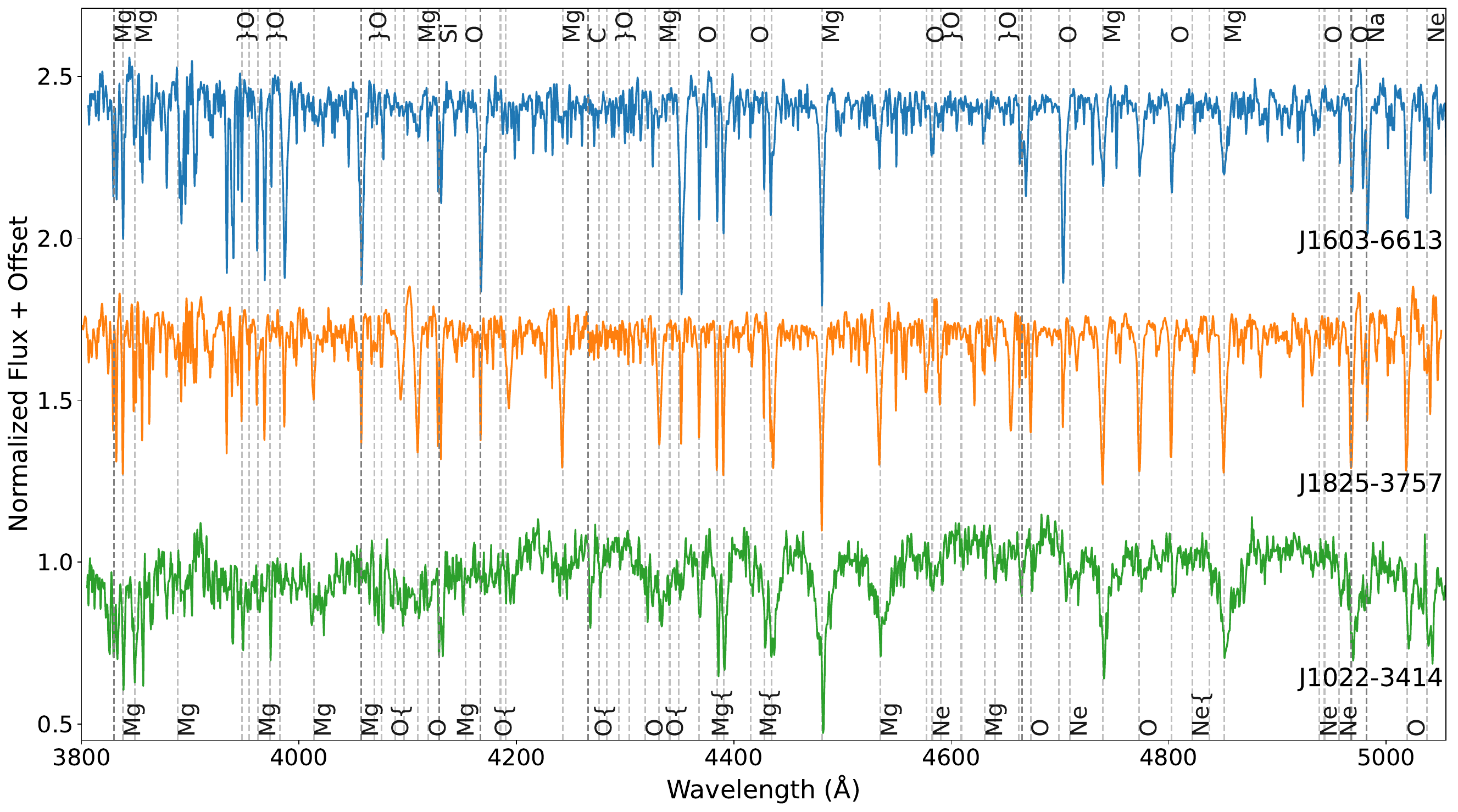}
    \caption{A comparison between the normalized observed spectrum of \gstar\ and two known \lp\ type stars from \citet{raddi2019} retrieved from the ESO archive. The two known \lp\ type stars have been smoothed to the same resolution as \gstar\ and wavelengths are at rest. Lines of magnesium, neon, and oxygen are marked. \gstar\ shows broader lines than the other two.}
    \label{fig:comp}
\end{figure*}

The target was observed in April 2025 and 3 consecutive exposures were taken to check for radial velocity (RV) variability. We used the 1200B+97 grism with a 0.7 arcsec slit and individual exposure times of $1500$\,s. We reduced the spectra with the FORS2 pipeline provided by the ESO reflex environment \citep{2013A&A...559A..96F}. We checked and applied standard data reduction procedures including bias, flat-field correction, and wavelength calibration. No significant radial velocity (RV) variability was detected within $10$~\kms.  We can therefore rule out short period binaries, but follow-up spectra will be needed to rule out long period binaries (orbital period of the order of days). The spectra were co-added after barycentric correction to achieve a higher signal-to-noise ratio (S/N) for the spectral analysis. The spectrum revealed that J1022-3414 was not a hot subdwarf. Instead, absorption lines of metals, in particular Mg, O, and Ne were detected. The spectrum resembles two known \lp\ type stars J1825-3757 and J1603-6613 \citep{raddi2017,raddi2019}. Fig.\,\ref{fig:comp}. shows a comparison of these stars.

\section{Spectroscopy}
\label{sect:spec}

The co-added spectrum has a median S/N of $\sim$45 allowing for the identification of many individual elemental lines. Blue-shifted lines of \Ion{O}{i}, \Ion{Mg}{ii}, and \Ion{Ne}{i} are very clearly identified in the spectrum. Due to the lower \teff\, of the \lp\ type stars from \citet{raddi2019}, \Ion{Ne}{i} lines were visible only at the red end of the optical range with the strongest line at 6402\,\AA\ -- outside the wavelength coverage of our spectrum. However, the presence of higher excitation lines in our bluer spectrum reflect the higher \teff\ ($19600\pm700$ K) of this object. We also detect many weaker lines arising from \Ion{C}{ii}, \Ion{O}{ii}, \Ion{Mg}{i}, \Ion{Al}{ii/iii}, \Ion{Si}{ii}, and \Ion{Ca}{ii}. The strongest and sharpest lines in the spectrum were compared with the model to infer a high barycentric radial velocity of $-390\pm20$ \kms, again similar to the class of \lp\ type objects..

The spectrum was fitted using the latest version of the Koester white dwarf model atmosphere code \citep{koester2010}.
While primarily designed for modelling white dwarfs, this LTE (local thermodynamic equilibrium) code has been successfully used
in the past for modelling \lp-stars \citep{raddi2018,raddi2019} and D$^6$ stars \citep{hollands2025}.
We included all detected elements in the model calculation as listed above,
but also all other elements detected in the \lp-like star J1825$-$3757 \citep{raddi2019},
at the same abundances relative to neon, which was the most abundant element for those stars.
Approximately 25\,000 lines spanning the EUV to the NIR were included in 
both the atmospheric structure calculations and the spectral synthesis.
It was necessary for this hotter star to include as many lines as practicable,
as half the flux was found to be emitted between 1500--3000\,\AA, and so the effect of metal line
blanketing in this region, and its effect on redistributing flux to redder wavelengths, needed to be considered.
Similarly to \citet{hollands2025} who modelled a $\approx16\,000$\,K D$^6$ star,
we found that the majority of these metal lines were very narrow (full width half maxima of order $0.01$\,\AA).
While these individual lines have a negligible impact on the continuum flux,
their combined contribution to the emergent flux could not be ignored,
and leads to an effective lowering of the optical continuum after convolving to the instrumental resolution.
For wavelengths bluer than $\approx 4500$\,\AA, the continuum opacity of our models is dominated by bound-free
opacities of \Ion{O}{i}, \Ion{Ne}{i}, and \Ion{Mg}{ii}. These bound-free opacities, as well as those from
other abundant elements in our models were included from the Opacity Project \citep{1993A&A...275L...5C}.
At redder wavelengths, the continuum is instead formed from the free-free opacity of these metals.
For all model calculations, fluxes were converged to 0.1 \% accuracy.

For the fit itself, we made \teff, \logg, and number abundances of C, O, Mg, Si, and Ca (relative to Ne)
free parameters, minimizing the $\chi^2$ between the data and the model (least-squares fit),
the latter of which was convolved to an instrumental resolution of 1.9\,\AA.
A fixed radial velocity of $-390$\,\kms\  (as measured above), was applied to all models throughout.
To account for the uncertain interstellar reddening and inaccuracies in flux-calibration,
at each step in the fit, we took the ratio of the spectrum and model, fitting with a 5th-order polynomial.
The model fluxes were then normalized by rescaling with this polynomial, leaving only features that varied rapidly with wavelength,
i.e.\ absorption lines. Since aluminium has only a small number of relatively weak lines, it does not have a strong effect on the atmospheric
structure. Therefore, we fitted the Al abundance independently by constructing a 1D grid of models,
reducing the number of free parameters in the former least squares fit.

\begin{figure*}
    \centering
    \includegraphics[width=0.95\linewidth]{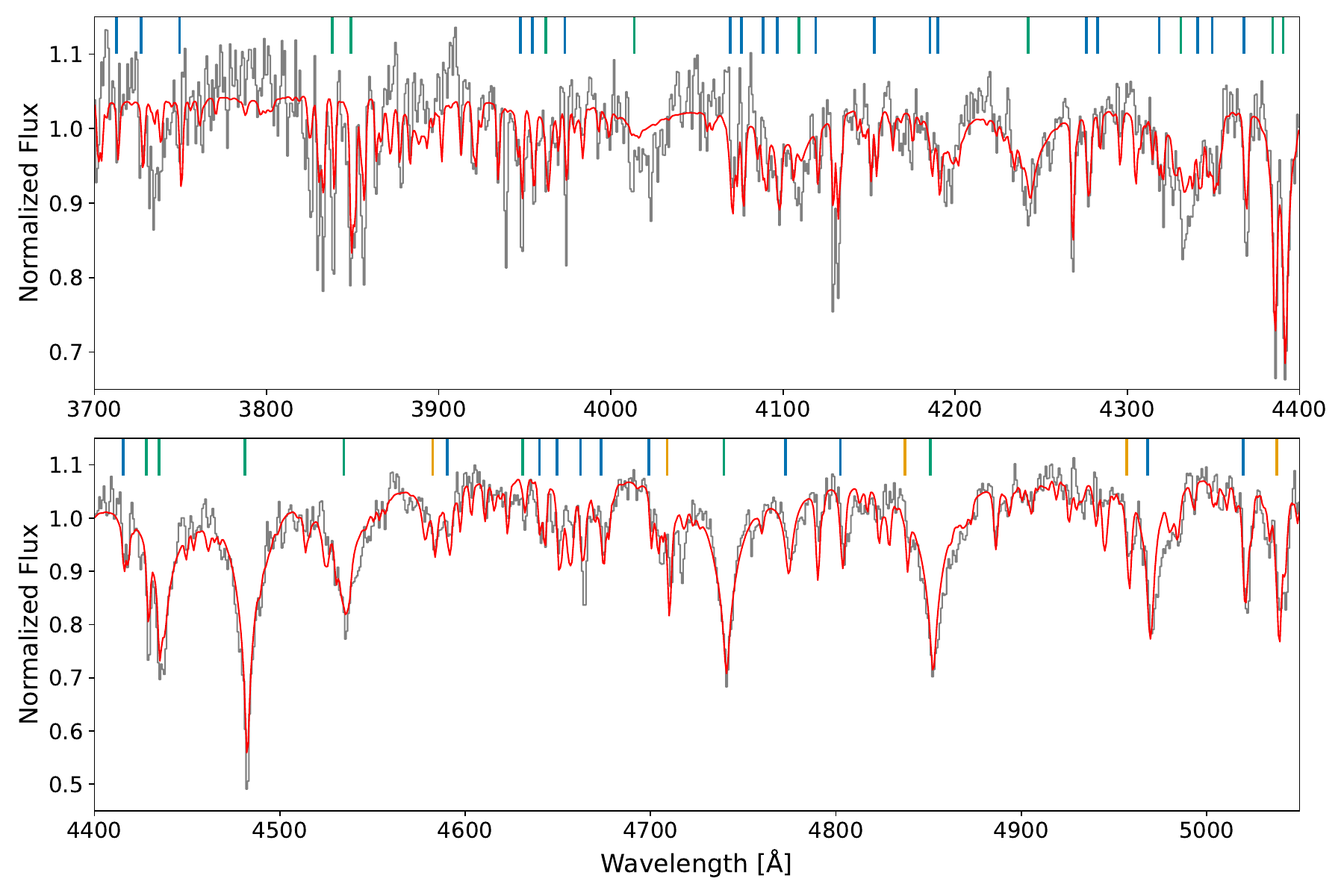}
    \caption{The best-fit spectrum in the rest-frame (red) overlaid on top of the co-added FORS2 spectrum (grey). Prominent lines of oxygen (blue), neon (orange), and magnesium (green) are marked. }
    \label{fig:spectrum}
\end{figure*}

The best-fitting model is illustrated in Fig.~\ref{fig:spectrum} and the corresponding atmospheric parameters and stellar abundances are listed in Table~\ref{tab:abundances}. Although neon is the most abundant element by number, abundances are reported
relative to oxygen, as this element was found to dominate the continuum opacity. The abundance of neon confirms the observed trend of neon-dominated atmospheres for \lp\ type stars, in stark contrast to D$^6$ stars which are carbon-dominated. 
Upper limits were derived for hydrogen, helium, and iron. Helium in particular was constrained better than previous studies owing to the higher temperature of this star. The upper limit of $\log(\rm He/O)=-2.8$ is lower than those of LP 40-365 and J1603-6613 as studied by \citet{raddi2019}. Since the O/Ne ratio of the best-fit model is extra-ordinary, two models which have abundances similar to ONe-cores of ultra-massive white dwarfs were also produced. These models did not fit the data as well, but are discussed in Appendix \ref{sec:ONemodels}.

\begin{figure}[h]
    \centering
    \includegraphics[width=0.5\textwidth]{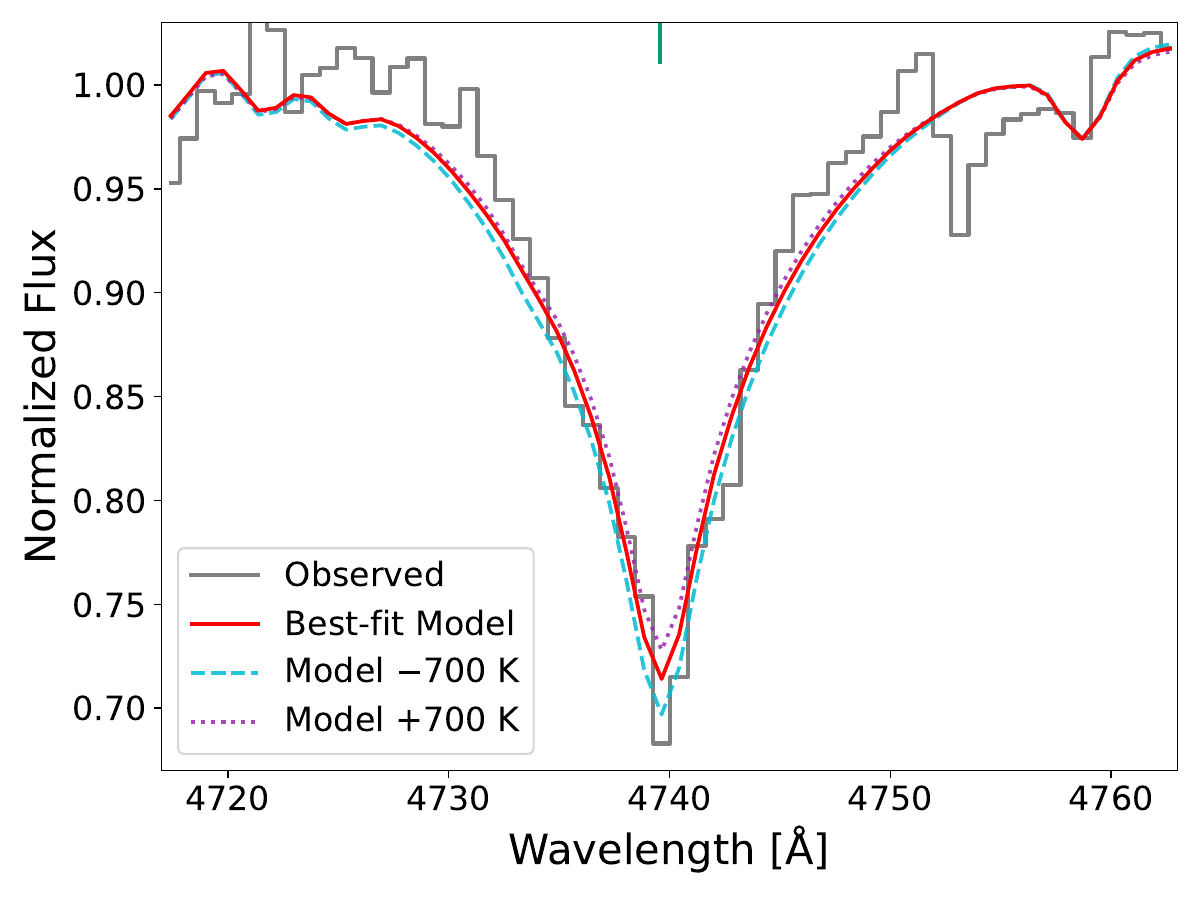}
    \caption{Comparison of the best-fit model (red) and the models that have \teff\, of $21300$ K and $18900$ K (in purple and cyan respectively), for a strong magnesium absorption line.}
    \label{fig:comp_teff}
\end{figure}

Similar to \citet{hollands2025}, the statistical uncertainties
were found to be small (e.g.\ $\sim$0.01\,dex for abundances), and so we estimated the uncertainties by constructing
one-dimensional grids of models for each parameter around the best
fitting values (steps of 50\,K in \teff\ and 0.05\,dex in \logg\ and abundances). These grids were visually compared to the data to determine the uncertainties. An example of the impact of the \teff\, on the Mg line at $4740$ \AA\, is shown in Fig.~\ref{fig:comp_teff}. The $\log g$ has a much higher uncertainty than normal white dwarfs due to the asymmetric impact on various metal lines. The Bayesian approach of \citet{2020NatAs...4..663H} was used to determine the 99th percentile upper limits of log(H/O), log(He/O), and log(Fe/O). However, spectra in the wavelength range $<3800$~\AA\ will be required to better probe Fe and Ni which are expected to form in Type Iax deflagrations.

Note that the uncertainties as described above do not account for uncertainties in atomic data, or in our models. While
the best-fitting model shown in Figure~\ref{fig:spectrum} shows generally good agreement, some discrepancies indicate
areas for improvement. For instance, there are a few lines that are present in our spectrum that are absent in our models.
Most notably lines with vacuum rest wavelengths of 3939\,\AA\ and 4717\,\AA. For the 3939\,\AA\ line, candidate lines within 1\,\AA\
of the line centre are \Ion{N}{iii}, \Ion{Ti}{iii}, and \Ion{Ni}{iii}. However, all candidate ions ought to have much stronger
lines at other wavelengths covered by our data which we do not observe. Similarly, for the 4717\,\AA\ line,
\Ion{Si}{iii} and \Ion{S}{ii} are also ruled out as candidates ions. In other cases lines are correctly included in our models
but fit poorly, with the 4129/4132\,\AA\ \Ion{Si}{ii} doublet the most notable example. While increasing the abundance in our
models will of course improve the fit to this doublet, it also introduces other \Ion{Si}{ii} lines that are not observed
in our data, and so the best-fitting abundance is a compromise between these considerations. This likely points to a need
for improved atomic data for some weaker lines that would not normally cause problems for other types of star.
It could also point to the fact that NLTE effects become more apparent at the hotter temperature of this object.

We are also limited by the number of lines we can feasibly include in our models. Each model requires several hours of computation time. When combined with the large number of necessary free-parameters, the least squares fit became computationally expensive.
As an experiment, we calculated a model with an order of magnitude more lines in the spectral synthesis,
finding a visually identical spectrum in the optical (after convolving to the instrumental resolution),
indicating that our reported abundances are likely to be accurate.
However we also found the GALEX UV flux reduced by about 0.1\,mag. This is comparable to the estimated error
on the GALEX photometry (see Sect.\ \ref{sect:sed}), and could correspond to a 2\% systematic uncertainty in
\teff, which is smaller than our reported uncertainty. 

Furthermore, the instrumental resolution limits the detection of any rotational broadening to $v\sin i>100$ \kms. We did not find any evidence of rotational velocity greater than 100 \kms. LP 40-365 has previously been shown to have rotational modulation in the light curve, but of the order $v\sin i<50$ \kms\, \citep{raddi2019,2021ApJ...914L...3H}.

Considering the fact that the atmosphere of the star is neither carbon nor oxygen dominated, our spectroscopic analysis does not favour a D$^6$ or violent merger scenario. A detailed kinematic study represents the logical next step to confirm the nature of the star. However, since the kinematic analysis can be better constrained by the inter-stellar reddening and the angular diameter of the star, we first describe the photometric analysis.

\begin{table}
\centering
\caption{Stellar atmospheric parameters and abundance ratios.}
\begin{tabular}{lcc}
\toprule
\toprule
Parameter & Value & Unit \\
\midrule
$T_{\mathrm{eff}}$ & $19600 \pm 700$ & K \\
$\log g$ & $5.1 \pm 0.4$ &  cgs\\
$v_{\mathrm{rad}} $& $-390 \pm 20$ & km\,s$^{-1}$ -\\
$\log(\mathrm{H/O})$ & $< -2.7$ & - \\
$\log(\mathrm{He/O})$ & $< -2.8$ & - \\
$\log(\mathrm{C/O})$ & $-2.2 \pm 0.3$ & - \\
$\log(\mathrm{Ne/O})$ & $0.30 \pm 0.15$ & - \\
$\log(\mathrm{Mg/O})$ & $-0.98 \pm 0.10$ & - \\
$\log(\mathrm{Al/O})$ & $-2.8 \pm 0.3$ & - \\
$\log(\mathrm{Si/O})$ & $-2.8 \pm 0.3$ & - \\
$\log(\mathrm{Ca/O})$ & $-4.7 \pm 0.3$ & - \\
$\log(\mathrm{Fe/O})$ & $< -3.3$ & - \\
\midrule
$\log(\Theta)$ & $-11.285 \pm 0.007$ & dex(rad) \\
$E(44-55)$ & $0.126 \pm 0.009$ & mag \\
\bottomrule
\end{tabular}
\label{tab:abundances}
\end{table}

\section{Photometry}
\label{sect:sed}

\begin{figure}
    \centering
    \includegraphics[width=0.95\linewidth]{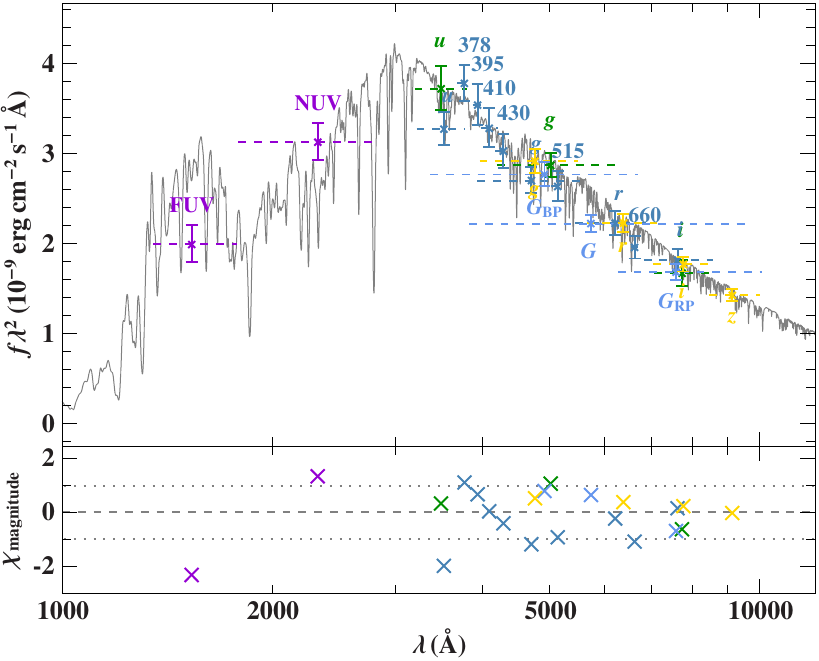}
    \caption{The SED of \gstar\,. Photometry was queried from DELVE DR2 \citep{2022ApJS..261...38D}, SPLUS DR4 \citep{2024A&A...689A.249H}, Gaia DR3 \citep{Gaia3}, SkyMapper \citep{skymapper}, and GALEX \citep{2005ApJ...619L...1M}. The model spectrum (in gray) is smoothed for visual clarity. The flux is multiplied with $\lambda^2$ to reduce the flux steepness. 
    }
    \label{fig:phot}
\end{figure}

We used the best-fit spectral model to model the stellar energy distribution. We fit the angular diameter ($\Theta$) and the reddening $E(44-55)$ of the star using the $\chi^2$-fitting method described in \citet{Uli2018}. We used the monochromatic interstellar extinction law from \citet{2019ApJ...886..108F}, with a fixed total-to-selective extinction parameter $R(55) = 3.02$, which is an average value representative of the diffuse Galactic interstellar medium. The SED fit along with the corresponding photometry is shown in Fig.~\ref{fig:phot}. 
We derive $E(44-55) = 0.126\pm0.009$\,mag, corresponding to $E(B-V) = 0.11\pm0.02$\,mag, which is only slightly higher than the line-of-sight value of $E(B-V) = 0.087$\,mag from \citet{2011ApJ...737..103S}. 
The photometric parameters are shown in the lower part of Table\,\ref{tab:abundances}. 

\section{Kinematic analysis}
\label{sect:kin}

The main limitation on modelling the kinematics of these stars is the large uncertainty on their \gaia\ parallaxes which limits our ability to constrain their distances. Their intrinsic absolute magnitudes are not well constrained, showing a spread of approximately $M_G\approx3.5-8$~mag. Nevertheless, once spectroscopic information is available, kinematics provide a key discriminant between stars originating from violent mergers, double detonations, or deflagrations. In particular, the expected ejection velocities occupy distinct regions of parameter space for the different progenitor channels. We model the kinematics for the case of \gstar, using the \gaia\ parameters listed in Table~\ref{tab:kinematics}. The parallax is corrected for the zero-point following \citet{Lindgren2021} using the python package \textsc{gaiadr3-zeropoint}.

\subsection{The lower limit of distance}

The $2\sigma$ \gaia\ upper limit on the parallax corresponds to a lower-limit on the distance of $1.8$ kpc. This corresponds to a tangential velocity ($V_{\mathrm{t}}$) of $252$ \kms. Combined with the radial velocity the heliocentric spatial velocity of the star would be $464$ \kms. Corrected for values of the Sun's motion from \citet{2010MNRAS.403.1829S}, the Galactocentric lower bound is $\sim 706$ \kms, suggesting that the star was boosted due to Galactic rotation. This velocity is higher than the local escape velocity $\sim550$ \kms\,\citep{2021A&A...649A.136K}, making the star unbound. A lower limit of the absolute magnitude of the star in the \gaia\ $G$-band ($
M_g = m_g - 5 \log_{10}\left(\frac{d}{10~\mathrm{pc}}\right) - A_g$) would be $7.45$ mag, close to the lower end of the observed \lp\ type stars \citep{kareemfast}.

\subsection{Distance sampling with flat magnitude prior}

We use the Python package \texttt{emcee} \citep{emcee-Foreman-Mackey-2013} to sample the posterior distribution of the Gaia $G$-band absolute magnitude $M_G$, the mass, and the proper motion components of the target. Similar to \citet{kareemfast} we rely on the \gaia\ $\varpi$, proper motions $(\mu_{\alpha*}, \mu_\delta)$, apparent magnitude $G$, and reddening $\rm E(B-V)$. We use the astrometric covariance matrix $\Sigma_\mathrm{obs}$ constructed from \gaia\ uncertainties and correlations. Following \citet{hollands2025}, we add spectro-photometric constraints: angular diameter $\log \Theta$ and surface gravity $\log g$. To make sure that the sampling is physical we add a mass prior: $M < 1.3\,M_\odot$, using $R = \frac{d \Theta}{2}$ and $M = \frac{g\,R^2}{G}$. The Likelihood function for vector $(\varpi, M_G)$ is compared using $\varpi_\mathrm{pred} = 10^{(M_G - G + 10 + A_G)/5} $, where $A_G$ is the extinction, $2.7\times E(B-V)$, from \citet{2018MNRAS.479L.102C}. We use the astrometric vector,
    \[
    \boldsymbol{\alpha}_\mathrm{pred} = (\varpi_\mathrm{pred}, \mu_{\alpha*}, \mu_\delta).
    \]
    and the astrometric log-likelihood:
    \[
    \ln \mathcal{L}_\mathrm{astro} = -\frac{1}{2} (\boldsymbol{\alpha}_\mathrm{pred} - \boldsymbol{\alpha}_\mathrm{obs})^T \Sigma_\mathrm{obs}^{-1} (\boldsymbol{\alpha}_\mathrm{pred} - \boldsymbol{\alpha}_\mathrm{obs}).
    \]
We add the predicted $\log g$ to the likelihood as:
 \[
\ln \mathcal{L}_{\log g} = -\frac{1}{2} \sum_i \frac{(\log g_{\rm pred} - {\log g}_{\rm obs})^2}{\sigma_{\log g}^2}.
\]
The uncertainties on $\Theta$, and $E(B-V)$ are added in quadrature to the likelihood for $\log g$ and $\varpi$ respectively.
     
The mass is initialized uniformly between $0.1-1.3$ \msun\ and the upper bound of ($<1.3$ \msun) is conservative, reflecting an upper limit of remnant masses which might be left behind after deflagration of a Chandrasekhar-mass white dwarf. Work on the D$^6$ stars \citep{Bhat1,glanz1,Bhat2,wong2025} shows that objects less massive than $0.1$ \msun\ or more massive than $0.5$ \msun\ will either be too cold or too hot respectively, compared to our star. Work on fully-convective and heated C/O and O/Ne white dwarf models shows the same result, since the Kelvin-Helmholtz cooling is set by the mass and luminosity \citep{ken2025}. We explore this prior in the Appendix \ref{sec:Massprior}.
We initialize the absolute magnitude by using a uniform distribution for $M_G$ based on the other known \lp\ type stars, assuming a flat prior on $M_G$ between 0 and 15, and a positive prior on $\varpi$. We also apply an upper bound of $20$ $\rm kpc$ on the derived distance, which corresponds to an upper bound of $\sim$2500 \kms\ on the ejection velocity of the star. $2500$ \kms\ reflects a conservative bound which is achievable only in rare cases through the violent merger scenario of \citet{pakmor2025} or for white dwarfs more massive than $1.1$ \msun\ in the D$^6$ scenario \citep{kareemfast}. 

The total log-posterior is then:
\[
\ln P(x \mid \mathrm{data}) = \ln \mathcal{L}_\mathrm{astro} + \ln \mathcal{L}_\mathrm{x} + \ln P_\mathrm{prior}.
\]

We use $128$ walkers for $10000$ steps and discard the first $4000$ as burn-in. This allows us to reach chain-lengths greater than $50$ auto-correlation times, recommended for convergence, and have enough independent samples after burn-in.

\begin{figure*}
    \centering
    \includegraphics[width=0.95\linewidth]{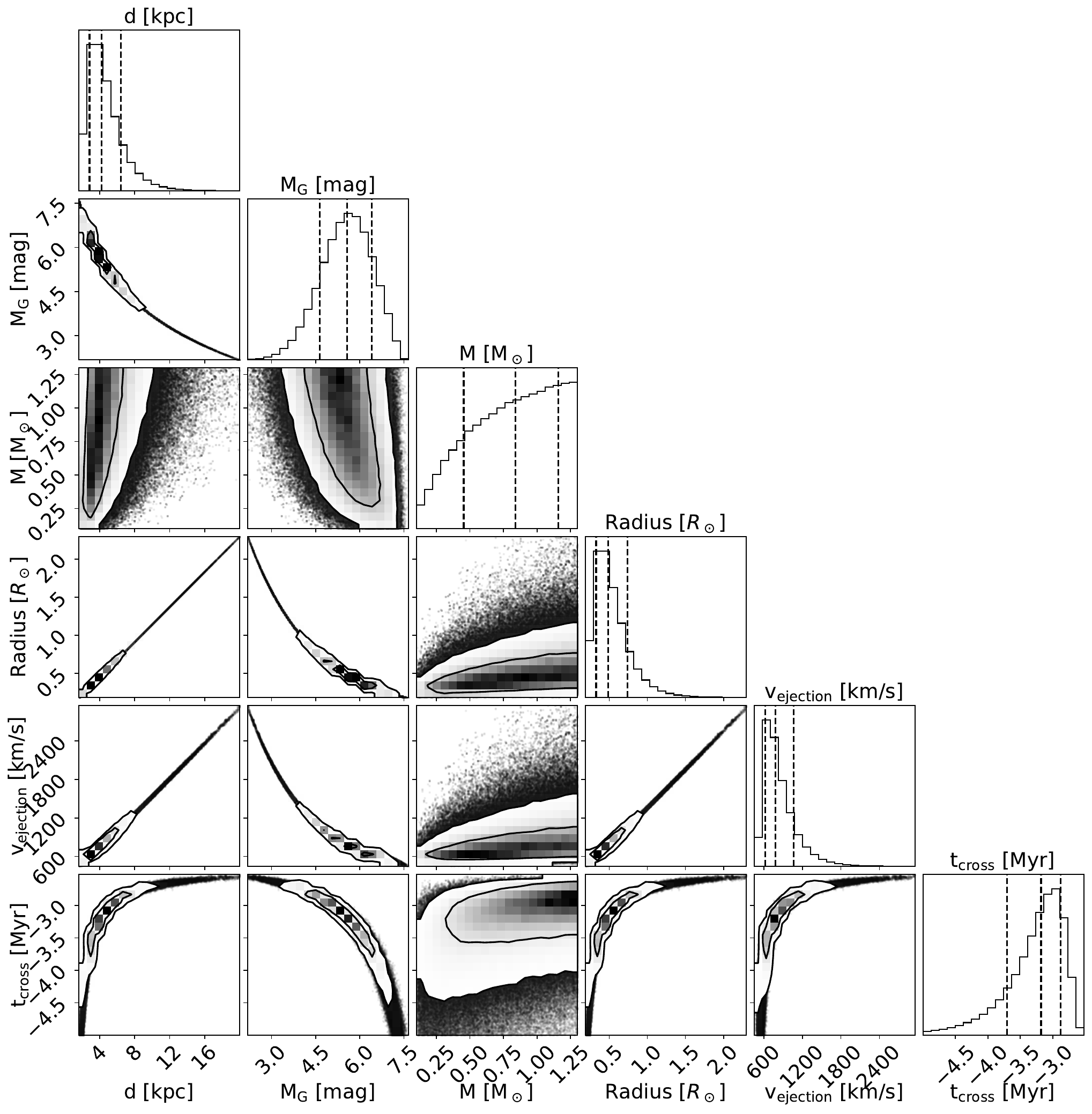}
    \caption{Posterior distributions of the distance, absolute magnitude, mass, radius, ejection velocity, and midplane-crossing time. Vertical dashed lines represent $16-84\%$ confidence intervals.}
    \label{fig:corner1}
\end{figure*}

\begin{table}
\centering
\caption{Summary of parameters sampled with emcee with mode, median, and 16--84\% confidence intervals.}
\label{tab:stats}
\begin{tabular}{lccc}
\toprule
\toprule
Parameters & Mode & Median & 68\% CI \\
\midrule
Distance [kpc] & 3.49 & 4.17 & (2.79, 6.41) \\
$M_\mathrm{G}$ [mag] & 5.49 & 5.58 & (4.65, 6.45) \\
Mass [M$_\odot$] & 1.28 & 0.83 & (0.44, 1.16) \\
Radius [R$_\odot$] & 0.40 & 0.48 & (0.32, 0.74) \\
$v_\mathrm{ejection}$ [km/s] & 650 & 771 & (615, 1059) \\
$t_\mathrm{cross}$ [Myr] & -3.79 & -3.19 & (-3.75, -2.88) \\
\bottomrule
\end{tabular}
\end{table}

The corner plot is shown in Fig.~\ref{fig:corner1} and the corresponding statistics are given in Table~\ref{tab:stats}. The radius, ejection velocity, and time of flight are derived from the sampled parameters. The high correlation between the parameters is clear with the parallax being the main source of uncertainty. The Galactocentric ejection velocity is calculated assuming a straight line path and subtracting the rotation of the Galaxy at plane crossing, similar to \citet{kareemfast}. We assume a circular velocity of $240$ \kms. While there seems to be a positive correlation between mass and ejection velocity, this correlation appears to be small. The uncertainty in $\log g$ shows up in the mass distribution and without any other prior is very uncertain. Higher mass, higher velocity stars may be produced in the double detonation mechanism, but that scenario is excluded due to the spectroscopic results since carbon is not the dominant element.

A flat magnitude prior ($p(M)=\text{const}$) implies an inverse distance prior ($p(d)\propto d^{-1}$). From geometric considerations it can be shown that this implies $\rho(d)\propto d^{-3}$. For the case of ejected runaways this is most likely not true. Therefore, we also tested a constant density prior which implies $p(d)\propto d^2$, and should be valid for such runaway stars up to a distance scale proportional to their thermal times (which sets the timescale for their inflated radii when they become observable). This prior is discussed in Appendix A.

\subsection{Kinematic age}

We also calculate precise trajectories from the sampled distances and the measured radial velocity as described in \citet{andreas,2018Andreas}. We use a revised version of the mass model introduced by \citet{1991RMxAA..22..255A},  which is called Model 1 in \citet{andreas}, for the Galactic potential. We numerically integrate the equations of motion for the stars using a $4$th-order Runge-Kutta solver with adaptive step size. The local standard of rest velocities ($U$,$V$,$W$)$_\odot$ are $(11.10\pm1.25, 12.24\pm2.05,7.25\pm0.62)$ \kms, taken from \citet{2010MNRAS.403.1829S}.

\begin{figure}
    \centering
    \includegraphics[width=0.95\linewidth]{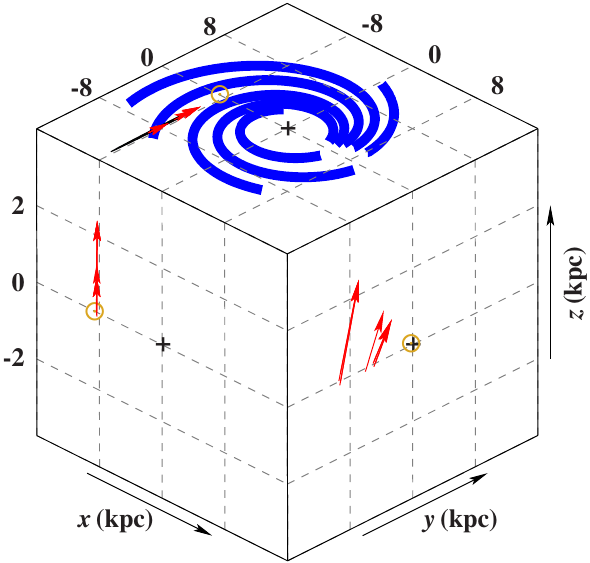}
    \includegraphics[width=0.95\linewidth]{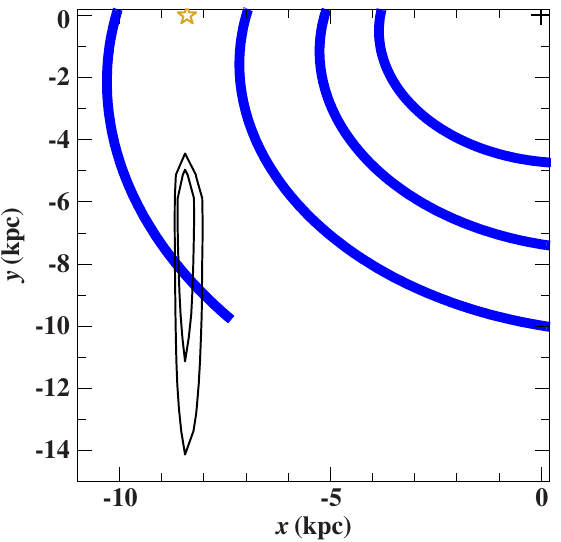}
    \caption{Top panel: 3D kinematic trajectory of the star since disk ejection. Blue curves show representative spiral arms from \citet{2014A&A...569A.125H}. The Sun is marked by the yellow circle and the Galactic centre by the black plus. Bottom panel: 2D plane-crossing contours marking regions where $68$\% and $95$\%  of the trajectories intersected the Galactic
plane when uncertainties were propagated. }
    \label{fig:kin}
\end{figure}

Out of the $10^5$ computed trajectories the star is bound for $0.005\%$ of the simulated trajectories. Nine representative 3D trajectories, corresponding to the $68\%$ confidence thresholds are shown in Fig.~\ref{fig:kin}. The star has a high uncertainty in the Galactic $y$ coordinate stemming from the distance uncertainty but was ejected almost perpendicular to the disk. The ejection velocity of the star calculated by removing the Galactic rotation at the point of disk crossing, and assuming a $1$ kpc disk scale height lies within $608-1190$ \kms\ ($16-84\%$ confidence interval). The corresponding time of flight of $2.74^{+0.34}_{-0.16}$ Myr is in line with those of other \lp\ type stars.

\begin{table}
\centering
\renewcommand{\arraystretch}{1.2}
\caption{Orbital parameters for \gstar.
Quoted uncertainties correspond to the 1$\sigma$ confidence intervals.}
\label{tab:kinematicparams}
\begin{tabular}{lclc}
\hline\hline
Parameter & Value & Parameter & Value \\
\hline
$X$ [kpc] &
$-8.31^{+0.07}_{-0.06}$ &
$X_f$ [kpc] &
$-8.37^{+0.06}_{-0.05}$ \\

$Y$ [kpc] &
$-3.52^{+0.77}_{-3.58}$ &
$Y_f$ [kpc] &
$-6.10^{+0.76}_{-3.56}$ \\

$Z$ [kpc] &
$1.32^{+1.16}_{-0.36}$ &
$Z_f$ [kpc] &
$0.00^{+0.10}_{-0.10}$ \\

$r$ [kpc] &
$9.09^{+1.89}_{-0.39}$ &
$r_f$ [kpc] &
$10.32^{+2.36}_{-0.47}$ \\

$V_X$ [km\,s$^{-1}$] &
$20.0^{+11.2}_{-6.7}$ &
$V_{X,f}$ [km\,s$^{-1}$] &
$7^{+16}_{-10}$ \\

$V_Y$ [km\,s$^{-1}$] &
$814^{+160}_{-58}$ &
$V_{Y,f}$ [km\,s$^{-1}$] &
$809^{+156}_{-62}$ \\

$V_Z$ [km\,s$^{-1}$] &
$406^{+468}_{-140}$ &
$V_{Z,f}$ [km\,s$^{-1}$] &
$411^{+462}_{-141}$ \\

$v_{\rm grf}$ [km\,s$^{-1}$] &
$872^{+410}_{-87}$ &
$v_{\rm grf,f}$ [km\,s$^{-1}$] &
$873^{+404}_{-96}$ \\

$v_{\rm grf}-v_{\rm esc}$ [km\,s$^{-1}$] &
$271^{+423}_{-99}$ &
$v_{\rm ej}$ [km\,s$^{-1}$] &
$716^{+474}_{-108}$ \\

Boundness [\%] &
0 &
$t_{\rm flight}$ [Myr] &
$2.74^{+0.35}_{-0.16}$ \\
\hline
\end{tabular}
\end{table}

\section{Evolutionary status and the problem with Type Iax survivors}
\label{sect:evol}

Partial deflagrations of CO white dwarfs have previously been studied as the progenitors of Type Iax SNe \citep{2022A&A...658A.179L}. Hybrid CONe white dwarfs and ONe white dwarfs are other systems which have been studied \citep{2015MNRAS.450.3045K}.  However, recent observations show that the surface abundances, mainly the high neon abundances of observed surviving remnants, are incompatible with the nucleosynthesis yields of such explosions from CO white dwarfs \citep{raddi2019}. Abundance patterns from deflagrations in ONe white dwarfs might be better suited to explain the observed spectra which are dominated by neon and oxygen, but still do not produce neon which is more abundant than oxygen. In fact, no known standard mechanism has so far been shown to produce neon which is this abundant. These stars are significantly different from the hottest D$^6$ stars. Helium-deficient stars also include post-AGB stars called PG1159 stars. However, both D$^6$ and PG1159 stars have high carbon and oxygen abundances and only marginal amounts of neon \citep{2024A&A...682A..42W}. We performed envelope calculations \citep[the procedure is described in detail in][]{2020A&A...635A.103K} by integrating the atmospheric model downwards. We found that neon is less buoyant than oxygen in the star’s outer convective zone and therefore, diffusion processes are unlikely to explain the overabundance of neon that we observe. Furthermore, the remnant masses and kicks received through these deflagration \citep[see for example][]{2022A&A...658A.179L} are too low to explain the observed velocities by at least a factor of 2 for the fastest simulated velocities. 

However, heated post-explosion models can still help shed some light on the evolutionary state of these stars especially to describe their inflated states \citep{2019ApJ...872...29Z}. Using the 1D stellar evolution code Modules for Experiments in Stellar Astrophysics (MESA, \citealt{Paxton2011,Paxton2013,Paxton2015,Paxton2018,Paxton2019,Jermyn2023})  \citet{ken2025} showed that the Kelvin-Helmholtz cooling of initially fully convective ONe stars could explain the inflated nature of the observed \lp\ type stars. 

\begin{figure}
    \centering
\includegraphics[width=0.95\linewidth]{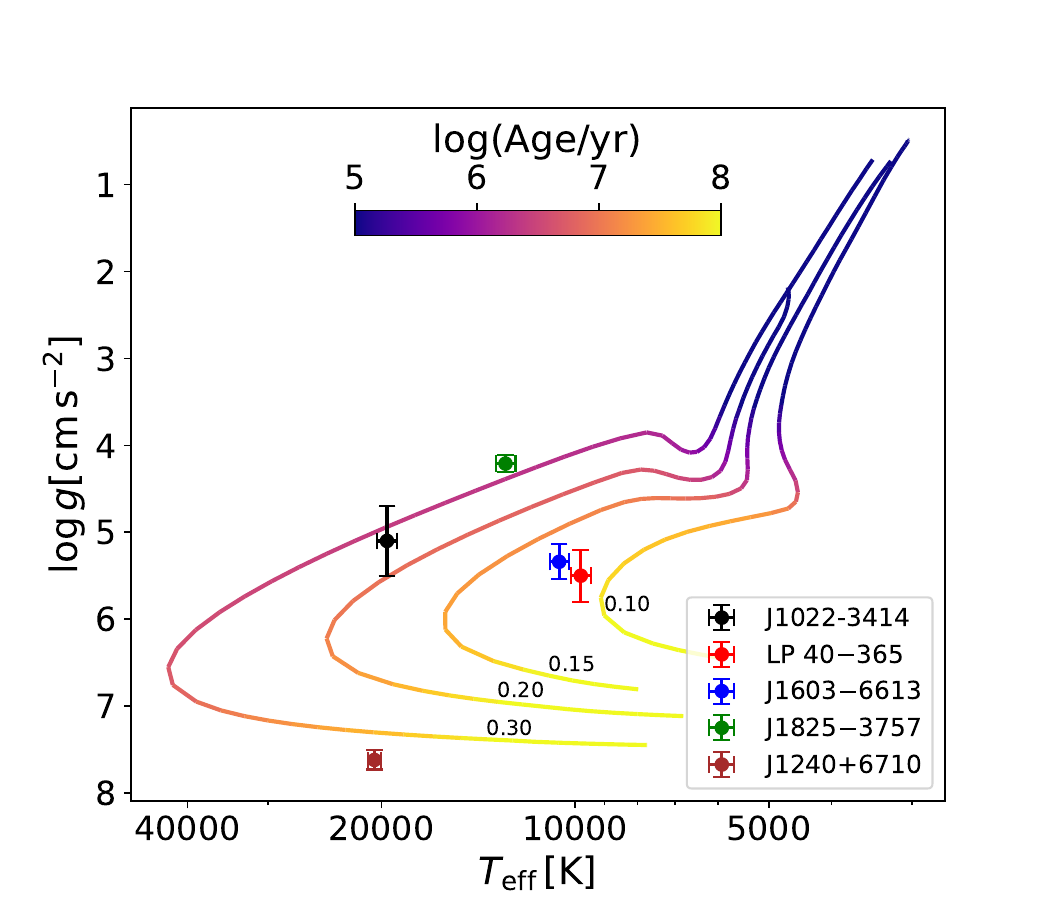}
    \caption{Kiel diagram of the observed stars plotted over evolutionary tracks of ONe dominated cooling convective balls from \citet{ken2025}. A lower limit of $0.20$ \msun\ may be established for the mass of \gstar\ due to the observed \teff.}. 
    \label{fig:evol}
\end{figure}

We show the Kiel diagram for the evolutionary tracks of \citet{ken2025} in Fig.~\ref{fig:evol}. We overplot a few known \lp\ type  stars taken from \citet{raddi2019} and the suspected \lp\ type star J1240+6710 \citep{boris2020,kepler2016} for which a spectral analysis with a corresponding $\log g$ exists. Our object is hotter than every other known \lp\ type star and more inflated than all but one. The lower $\log g$ shows this to be the case. The evolutionary track of the $0.3$ \msun\ model matches our observation quite well. The results of the kinematic analysis can not be compared further due to a lack of higher mass evolutionary models. While the ages of these models are prone to uncertainty due to starting entropy profiles \citep[see the discussion and comparison in][]{Bhat2} the upper limit of a few Myr matches the calculated kinematic age well. 

\begin{figure}
    \centering
    \includegraphics[width=0.95\linewidth]{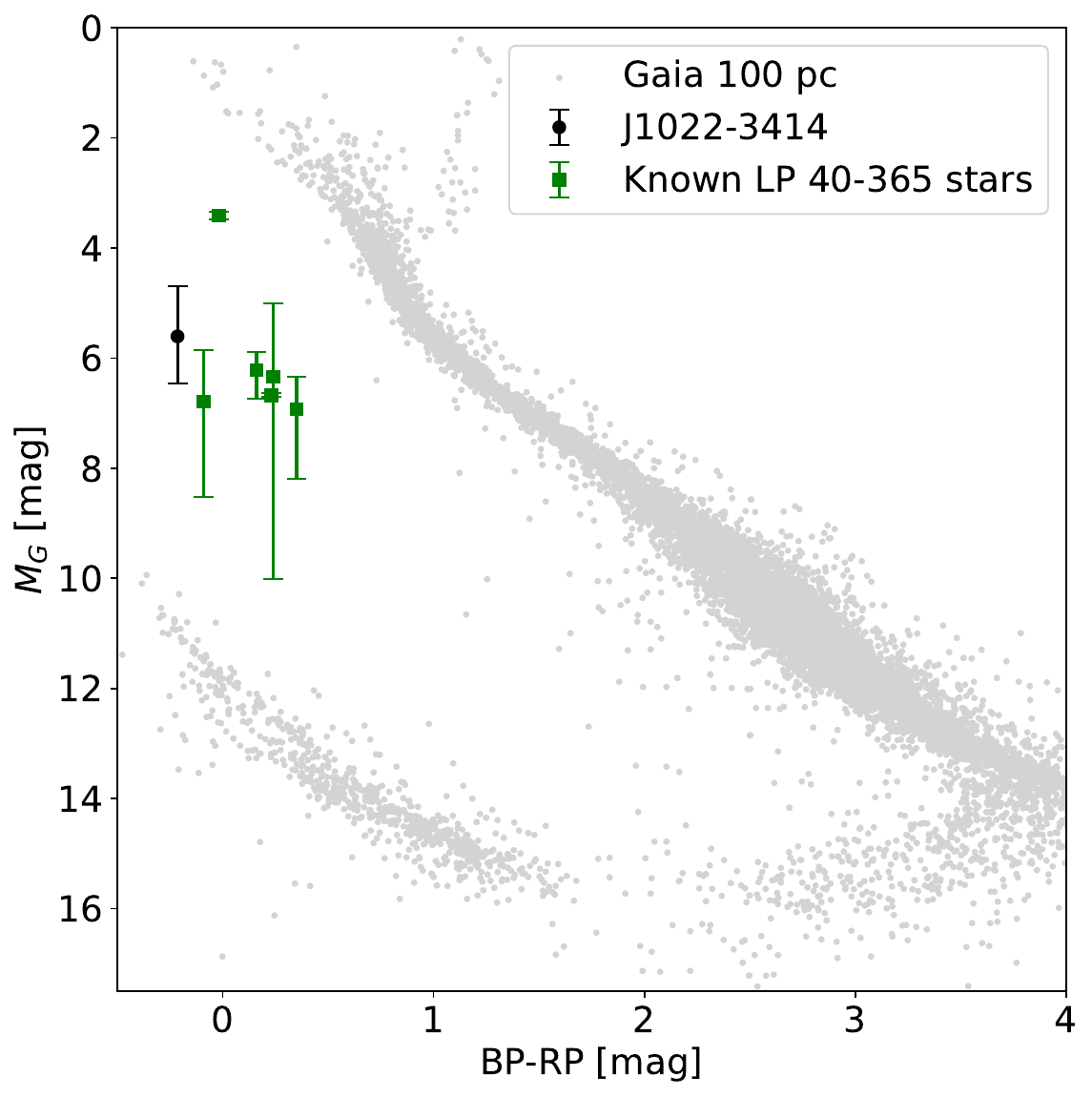}
    \caption{Colour-magnitude diagram of the confirmed \lp\ type stars from \citet{kareemfast} in green. Background stars (in grey) are selected from the \gaia\ 100 pc sample. \gstar\ is plotted in red with the absolute magnitude sampled using the emcee procedure described before.}
    \label{fig:cmd}
\end{figure}

The inflated nature of the star is also reflected in the colour-magnitude diagram, as shown in Fig.~\ref{fig:cmd}. Overplotted are the known \lp\ type stars taken from \citet{kareemfast}, and $10000$ stars within $100$ pc in \gaia, showing that the star is the hottest and second most inflated \lp\ type star known.

\begin{figure}
    \centering
    \includegraphics[width=0.99\linewidth]{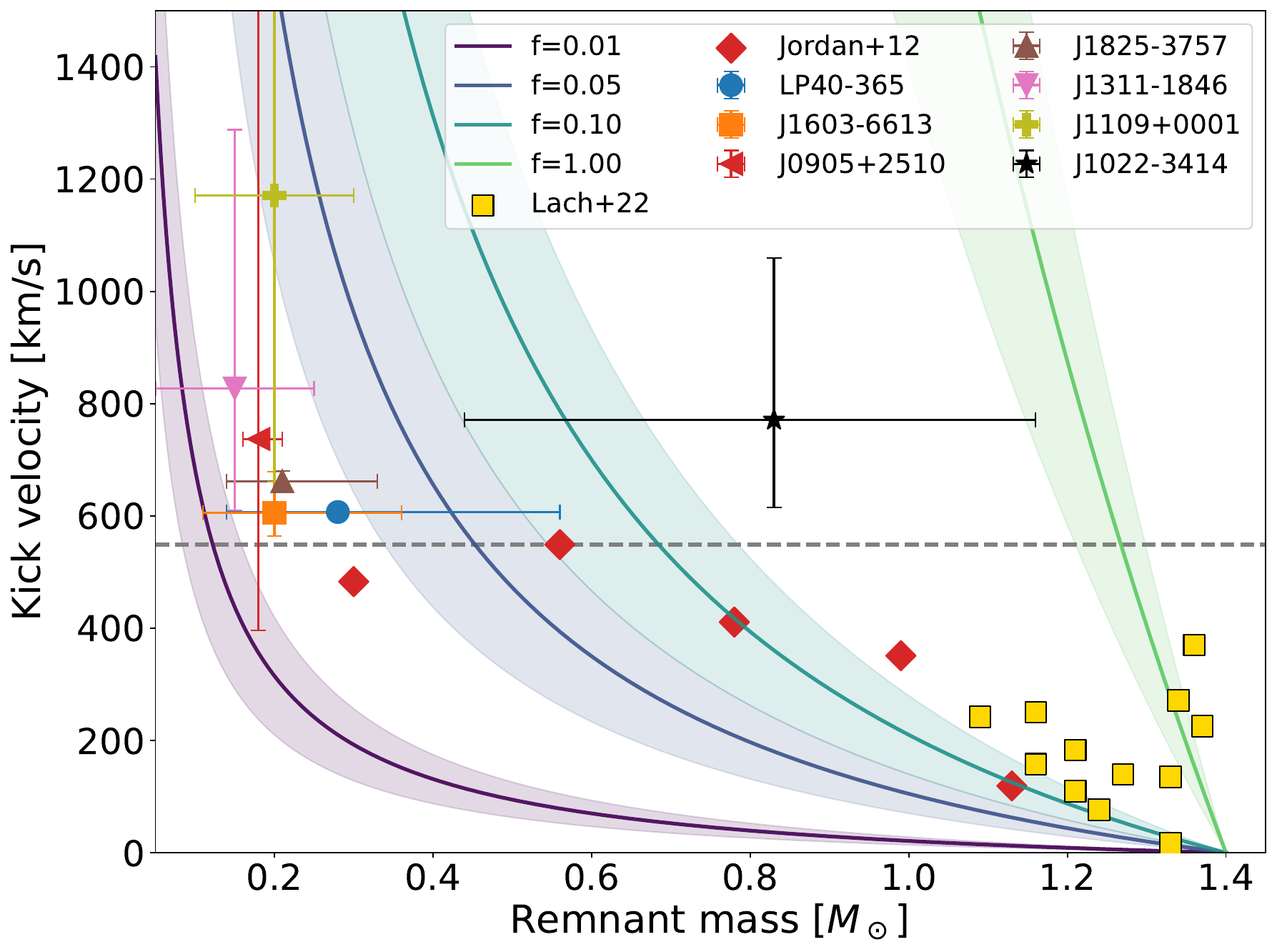}
    \caption{The ejection velocities of the observed \lp\ type stars as a function of their derived masses. Yellow squares and red diamonds are deflagration models from \citet{2022A&A...658A.179L} and \citet{2012ApJ...761L..23J}, respectively. Curves show different kick velocities from ejecta derived using Eq.1 for different asymmetry factors. The spread for each curve is computed using ejecta velocities between $3000$--$8000$\,\kms. The dashed grey line shows the maximum velocity achieved by any model so far.}
    \label{fig:vel}
\end{figure}

In Fig.~\ref{fig:vel} we plot the measured ejection velocities of the observed stars as a function of their mass. Along with the observations we also plot results of 3D simulations from the works of \citet{2012ApJ...761L..23J} and \citet{2022A&A...658A.179L}. While other models exist, in particular the oxygen deflagration models of \citet{jones2016} and \citet{2019A&A...622A..74J} which lead to lower mass remnants ($0.366$ \msun\ at the lowest), kick velocities are not reported for them. Similarly the models of \citet{2014MNRAS.438.1762F} are not plotted due to negligible kick velocities. We also plot curves for the expectation from a simple momentum conservation argument, wherein asymmetric ejecta with ejecta velocity $V_{\rm ej}$ with asymmetry factor $f$ from an initial white dwarf ($M_{i}$) lead to a remnant mass $M_{\rm rem}$ such that the kick velocity may be written as:
\begin{equation}
    V_{\textrm{kick}} = f\frac{(M_{\mathrm{i}}-M_{\mathrm{rem}})}{M_{\mathrm{rem}}}V_{\rm ej}
\end{equation}
Since it is assumed that the deflagrations are Chandrasekhar-mass, we assume the initial white dwarf mass to be $1.4$ \msun. The analytical expression matches well for $1-5$\% asymmetry, which is characterized by a slightly off-center ignition.

The simulated models so far cannot consistently produce such low-mass remnants with similar ejection velocities to those observed. Furthermore, these models have differences in their gravity solvers which rely on monopole or multipole expansion and which lead to different kick velocities. As such the higher velocities of the remnants from \citet{2012ApJ...761L..23J} might be a consequence of a different solver. \citet{2014MNRAS.438.1762F} pointed this out as they were unable to reproduce the kick velocities most likely because of their monopole solver. Shifting to an FFT-based solver, however, still did not significantly increase the kicks in their study. 

\section{Discussion}
\label{sec:discussion}
In this paper, we have detailed the spectroscopic and kinematic analysis of the hottest (\teff\ = $19600 \pm 700$ K) and second-most inflated (radius of $0.32-0.74$ \rsun) confirmed \lp\ type star. The spectroscopic results show a high neon abundance, followed by oxygen, in line with previous results of \lp\ type stars, and different from known runaway donors of Type Ia SNe which are carbon dominated. A Bayesian analysis using the \gaia\ astrometric and photometric parameters, the spectroscopic $\log g$ and the photometric angular diameter and reddening, suggests that the star is unbound from the Milky Way and was ejected $\sim2.8$ Myr ago from the Galactic disk. 

Taken together, the observations of all \lp, stars offer a comprehensive view of this class and pose a significant challenge to existing theoretical models. The following summarises the problems.

\begin{enumerate}
    \item \textit{The abundance issue}. While the presence of heavier elements has been suggested to be a signature of Chandrasekhar-mass deflagrations, no current deflagration model is able to reproduce the neon-rich abundances of the observed stars. Models only predict carbon- or oxygen-dominated survivors. Our spectroscopic models with higher O/Ne ratios are discussed in Appendix \ref{sec:ONemodels}. The data prefers models which have a lower O/Ne ratio. This seems to be the case for all the \lp\, type stars whose spectra have been modelled. The influence of reaction rates on the core mass fractions of ONe white dwarfs provides an interesting avenue to study. Similar studies have found that the carbon and oxygen fractions can change significantly (up to $60\%$) with reaction rates for CO white dwarfs \citep{2016ApJ...823...46F}. Similar results have been found for the mass compositions of heavier elements for core-collapse progenitors \citep{2018ApJS..234...19F}. The reaction rates would need to be significantly different than what is assumed in stellar evolution codes for the abundances to be reproduced, and would therefore have massive consequences for other areas of study as well.
    
    \item \textit{The ejection velocity issue.} From observations, slightly asymmetric, off-center ignited deflagrations which eject more than $0.9$ \msun\ and lead to bound remnants less than $0.5$ \msun\ are favoured. However,  most models show higher remnant masses ($>0.5$ \msun) and lower kicks. The orbital velocity of massive ONe accretors at the time of Roche-lobe overflow from a He star companion should be on the order of $200-300$ \kms. It should be even lower for main-sequence companions. In the highly improbable case that every deflagration gives a preferential kick in the direction of orbital motion, the minimal kick based on observed ejection velocities would have to be of the order $400$ \kms, significantly larger than what simulations can currently produce.
    \item \textit{The mass issue.} Similarly to ejection velocities, most models end up creating massive survivors ($>0.5$\msun). 
    These are far off from the $0.1-0.3$ \msun\ values inferred for most of the \lp\ type stars. For the case of \gstar\ the mass is not well constrained and a higher value is favoured, but so far this remains an exception.
\end{enumerate} 

Finally, an observational lack of fast ($>800$ \kms) He star donors (in comparison with accretors) to this channel further contributes to an already confusing picture. The unbound He-sdO US708 is the only known star of this kind \citep{hirsch2005,geier2015}, but is expected to be the surviving donor of a Type Ia supernova. A fast-moving ($\sim600-800$ \kms) population of He star donors (some of which may be partially degenerate) has been predicted for Type Ia SNe \citep{2025arXiv251111998R}. While matching the kick velocities observed for \lp\ type stars, they cannot explain the high observed neon abundances. This lack of observed He star donors may be alleviated slightly by a slower moving main-sequence star donor population. For now, \lp\ type stars present an opportunity to better understand some of the most peculiar supernovae in our Galaxy.

\begin{acknowledgements}
    We thank the anonymous referee for their helpful comments. A.B. was supported by the Deutsche Forschungsgemeinschaft (DFG) through grant GE2506/18-1. Based on observations collected at the European Southern Observatory under ESO programme 115.28G1.002. We thank Abinaya Swaruba Rajamuthukumar and Evan Bauer for insightful discussions. This work made use of the following software packages: \texttt{matplotlib} \citep{Hunter:2007}, \texttt{numpy} \citep{numpy}, \texttt{python} \citep{python}, \texttt{Cython} \citep{cython:2011}, \texttt{emcee} \citep{emcee-Foreman-Mackey-2013,emcee_10996751}, \texttt{corner.py} \citep{corner-Foreman-Mackey-2016,corner.py_14209694}, and \texttt{h5py} \citep{collette_python_hdf5_2014,h5py_7560547}.

Software citation information aggregated using \texttt{\href{https://www.tomwagg.com/software-citation-station/}{The Software Citation Station}} \citep{software-citation-station-paper,software-citation-station-zenodo}.
This work presents results from the European Space Agency (ESA) space mission Gaia. Gaia data are being processed by the Gaia Data Processing and Analysis Consortium (DPAC). Funding for the DPAC is provided by national institutions, in particular the institutions participating in the Gaia MultiLateral Agreement (MLA). The Gaia mission website is https://www.cosmos.esa.int/gaia. The Gaia archive website is https://archives.esac.esa.int/gaia.
\end{acknowledgements}

\bibliographystyle{aa}
\bibliography{references}

\begin{appendix}
\section{Distance prior}

The results for the distance prior are given in Table~\ref{tab:stats2}. The star is farther away and has a higher predicted mass. The ejection velocity of the star increases slightly to $742-1477$ \kms\, which is a consequence of the larger distance. The star is more inflated, with radius between $0.45-1.09$ \rsun. With this distance prior, the star is even more of an outlier when considering theoretical models, as discussed in Section \ref{sec:discussion}.

\begin{table}[ht]
\centering
\caption{Summary of parameters sampled with \texttt{emcee} with mode, median, and 16--84\% confidence intervals.}
\label{tab:stats2}
\begin{tabular}{lccc}
\toprule
Parameter & Mode & Median & 16--84\% CI \\
\midrule
Distance [kpc] & 4.86 & 6.06 & (3.93, 9.50) \\
$M_\mathrm{G}$ [mag] & 4.91 & 4.77 & (3.79, 5.71) \\
Mass [M$_\odot$] & 1.29 & 0.95 & (0.58, 1.20) \\
Radius [R$_\odot$] & 0.56 & 0.70 & (0.45, 1.09) \\
$v_\mathrm{ejection}$ [km/s] & 842 & 1013 & (742, 1477) \\
$t_\mathrm{cross}$ [Myr] & -2.81 & -2.91 & (-3.25, -2.72) \\
\bottomrule
\end{tabular}
\end{table}

\begin{figure}[ht]
    \centering
    \includegraphics[width=0.95\linewidth]{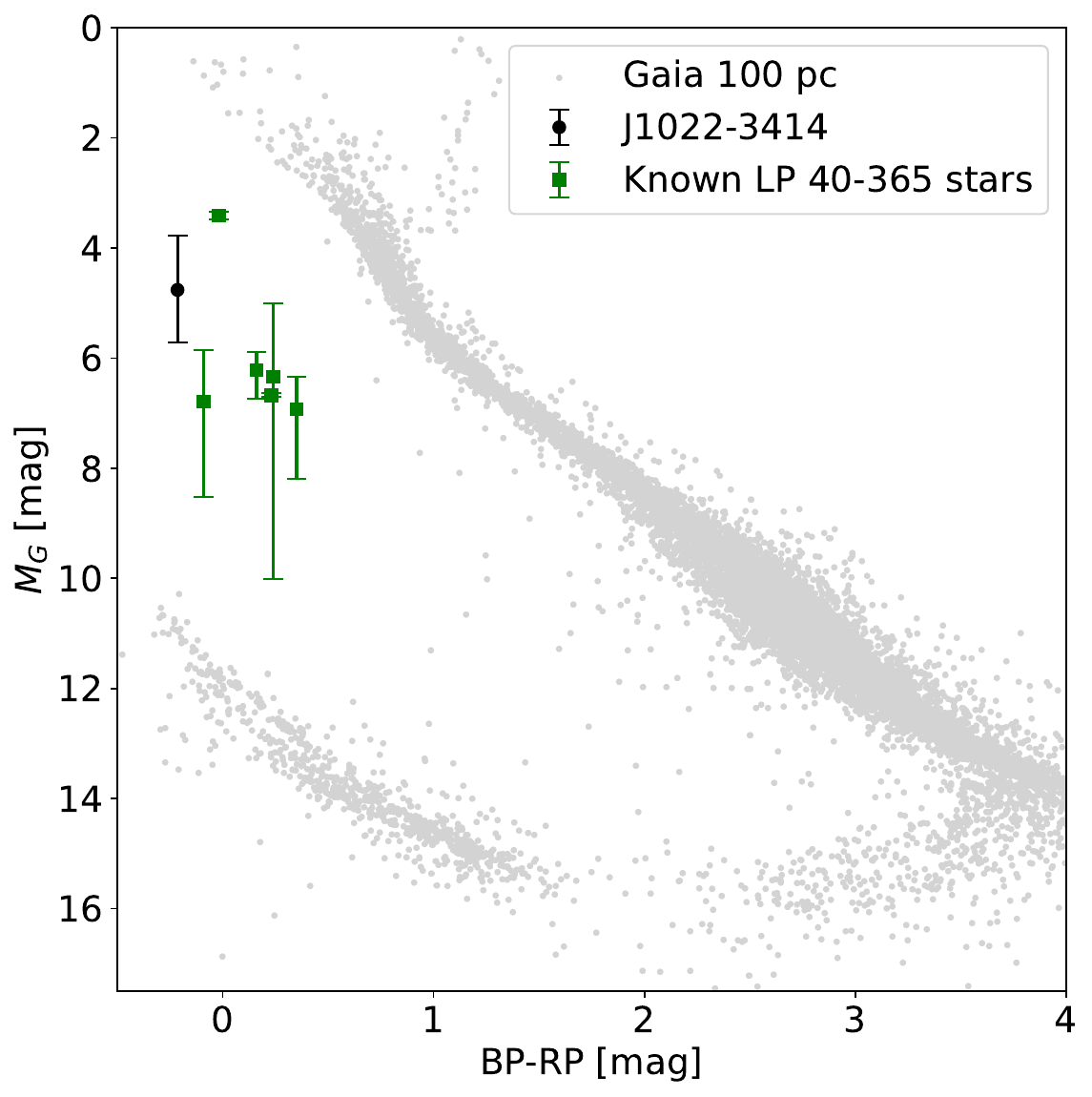}
    \caption{Same as Fig.~\ref{fig:cmd} but with a distance prior.}
    \label{fig:cmd2}
\end{figure}

\section{Priors on mass and their influence on kinematics}
\label{sec:Massprior}

\begin{figure}[ht]
    \centering
    \includegraphics[width=0.95\linewidth]{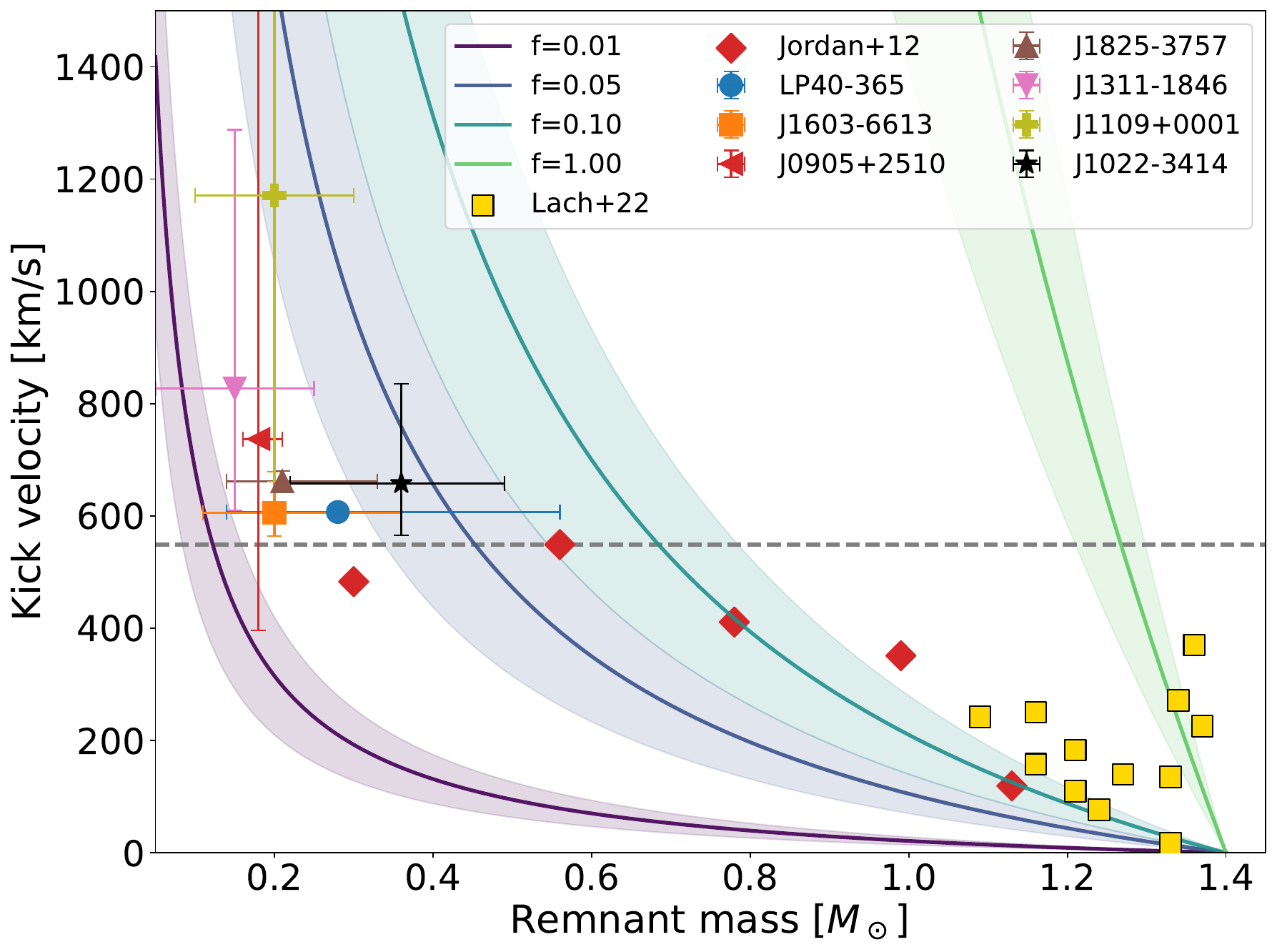}
    \caption{Same as Fig.~\ref{fig:vel} but for bounded mass priors between $0.1-0.5$ \msun.}
    \label{fig:ejectionvel2}
\end{figure}

\begin{figure*}
    \centering
    \includegraphics[width=0.95\linewidth]{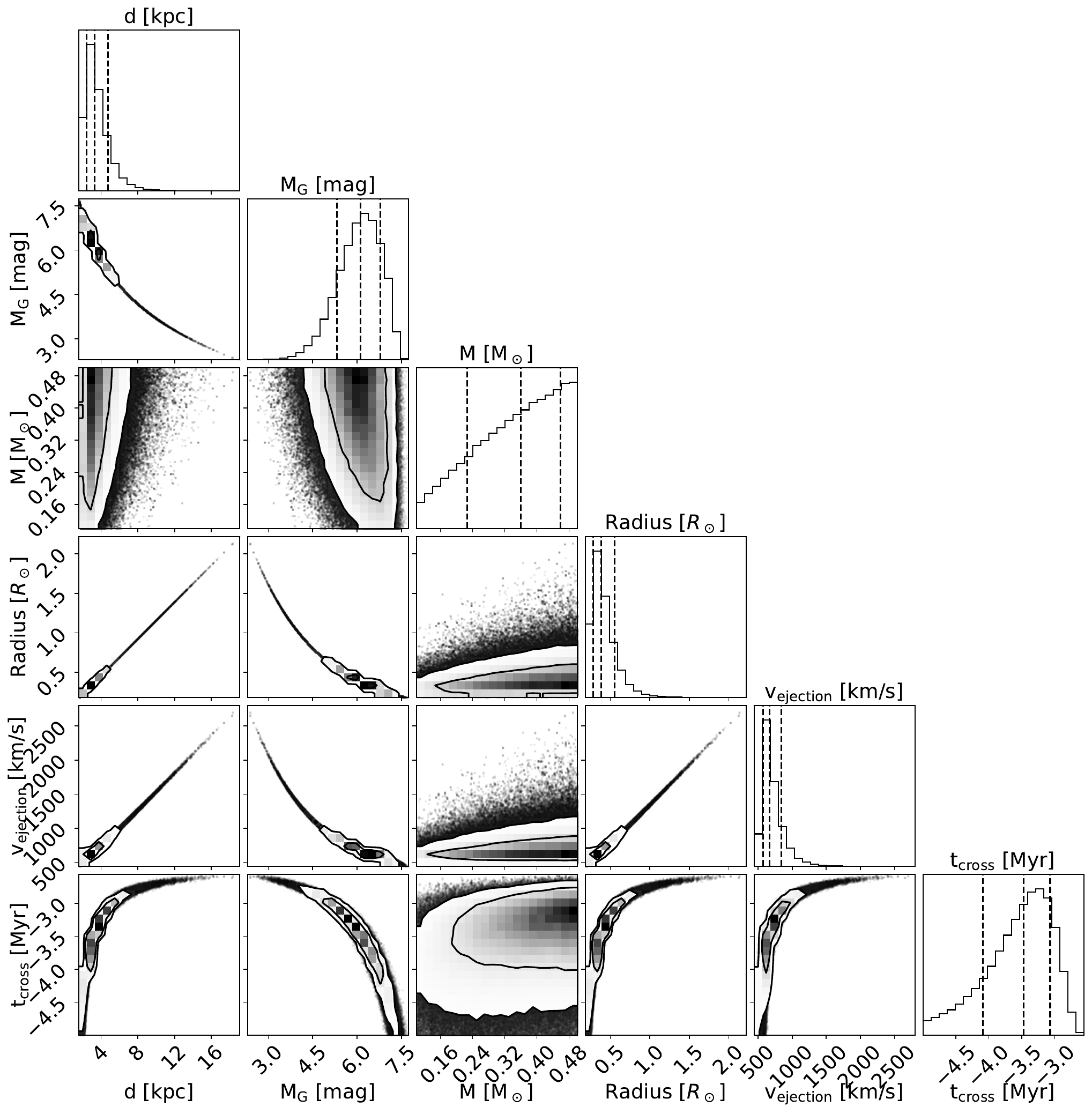}
    \caption{Same as Fig.~\ref{fig:corner1} but for a bounded mass prior.}
    \label{fig:corner2}
\end{figure*}

The spectroscopic \teff\ combined with previous studies of such stars allows us to place a conservative limit on the mass of the star which is between $0.1-0.5$ \msun. Using this prior, we can constrain the ejection velocity to be within $585-976$ \kms\ (68\,\% confidence). The corresponding values are given in Table~\ref{tab:stats3} and the corner plot is shown in Fig.~\ref{fig:corner2}. Since the mass is also constraining for the distance, there is only a $2-5\%$ difference in the flat magnitude prior and the distance prior for this case.
The updated velocity-mass diagram is shown in Fig.~\ref{fig:ejectionvel2}.

The posterior distributions of our mass samples indicate that a substantial fraction of the samples accumulate near the imposed upper limits. For the prior $M<1.3$ \msun, approximately $20\%$ of the posterior samples lie in the range $1.2$–$1.3$ \msun. Similarly, for the prior $M<0.5$ \msun, about $36\%$ of the samples fall between $0.4$ and $0.5$ \msun. This tendency toward higher masses is driven by the large uncertainties in \logg\ and the parallax. Nevertheless, this behavior does not affect our interpretation of the star’s ejection mechanism, as the inferred ejection velocity exceeds $600$ \kms\ in all cases. A positive correlation between mass and distance is present due to the larger radii at higher masses, but this effect is weak. Consequently, both the flight time and the ejection velocity remain largely unchanged.  

\begin{table}[ht]
\centering
\caption{Summary of stellar parameters with mode, median, and 16--84\% confidence intervals, assuming a $<0.5$ \msun\, prior on mass.}
\label{tab:stats3}

\begin{tabular}{lccc}
\toprule
Parameter & Mode & Median & 16--84\% CI \\
\midrule
Distance [kpc]               & 2.66 & 3.19 & (2.27, 4.67) \\
$M_\mathrm{G}$               & 6.32 & 6.18 & (5.35, 6.92) \\
Mass [M$_\odot$]             & 0.50 & 0.36 & (0.22, 0.46) \\
$\log g$                     & 4.81 & 4.80 & (4.48, 5.10) \\
Radius [R$_\odot$]           & 0.31 & 0.37 & (0.26, 0.54) \\
$v_\mathrm{ejection}$ [km/s] & 589 & 658 & (565, 835) \\
$t_\mathrm{cross}$ [Myr]     & -3.03 & -3.51 & (-4.27, -3.08) \\
\bottomrule
\end{tabular}
\end{table}

\section{Standard O/Ne-core models}
\label{sec:ONemodels}

\begin{figure*}
    \centering
    \includegraphics[width=0.95\linewidth]{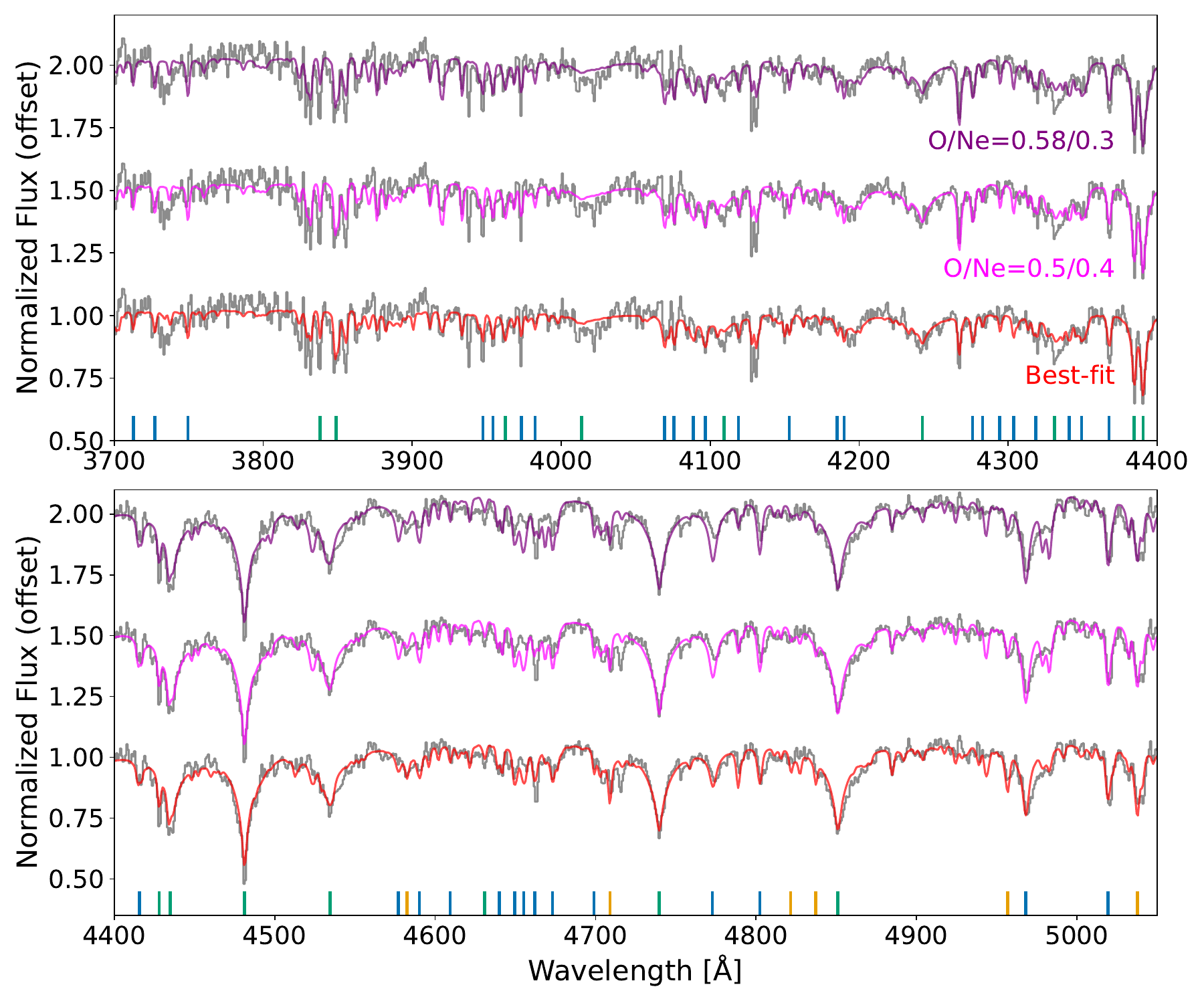}
    \caption{A comparison of three different models with different amounts of O/Ne. O/Ne ratios are given in terms of mass-fraction. The best-fit model has a nmass-fraction ratio of O/Ne of 0.28/0.67. The higher oxygen abundances lead to stronger oxygen lines such as the one at $4780 \,\AA$.}
    \label{fig:comp_models}
\end{figure*}

We also created models for two compositions with mass-fractions  $\mathrm{O/Ne}=0.5/0.4$ and $\mathrm{O/Ne}=0.58/0.3$ to reflect two possible core compositions of ONe white dwarfs. For the first model C/Na/Mg fractions were 0.01/0.03/0.06, and heavier metals were fixed to their solar mass fractions. This first composition with $\mathrm{O/Ne}=0.5/0.4$ reflects an extreme case of composition in standard ONe white dwarfs. The latter composition of $\mathrm{O/Ne}=0.58/0.3$ is taken from the abundance of $1.23$ \msun\, ONe white dwarf from \citet{2019A&A...625A..87C}. The comparisons of the models to the data are shown in Fig.\,\ref{fig:comp_models}. The $\chi^2$ of the models were 6014.43 for the best-fit neon-rich model, 6870.98 for O/Ne= $0.5/0.4$, and 7948.05 for O/Ne$=0.58/0.3$. This difference arises mainly due to the oxygen lines, which become a bit stronger and do not match the data as well. This can be seen for lines at $3750$ \AA, $4577$ \AA, $4772$ \AA, and $4980$ \AA. Therefore, our modelling of \gstar\, as with the other analysed \lp\, type stars, supports a Ne-enhanced composition compared to evolutionary models of ONe core white dwarfs.

\end{appendix}
\end{document}